\documentclass[superscriptaddress,twocolumn,nofootinbib,pra]{revtex4-2}

\usepackage{amssymb,xcolor}
\usepackage{amsmath}
\usepackage{amsbsy}
\usepackage{amsthm}
\usepackage{graphicx}
\usepackage{amsfonts}
\usepackage{epstopdf}
\usepackage{times}
\usepackage{txfonts}
\usepackage{hyperref}

\newcommand{\beq}{\begin{equation}}
\newcommand{\eeq}{\end{equation}}

\newcommand{\id}{\leavevmode\hbox{\small1\normalsize\kern-.33em1}}

\newcommand{\eq}[1]{Eq. (\ref{#1})}

\begin{document}

\title{Ghost imaging as loss estimation: Quantum versus classical schemes}

\author{A. Chiuri}
\affiliation{ENEA - Centro Ricerche Frascati, via E. Fermi 45, 00044 Frascati, Italy}
\author{I. Gianani}
\affiliation{Dipartimento di Scienze, Universit\'a degli Studi Roma Tre, Via della Vasca Navale 84, 00146, Rome, Italy}
\author{V. Cimini}
\affiliation{Dipartimento di Scienze, Universit\'a degli Studi Roma Tre, Via della Vasca Navale 84, 00146, Rome, Italy}
\affiliation{Dipartimento di Fisica, Sapienza {Universit\`a} di Roma, Piazzale Aldo Moro 5, I-00185 Roma, Italy}
\author{L. De Dominicis}
\affiliation{ENEA - Centro Ricerche Frascati, via E. Fermi 45, 00044 Frascati, Italy}
\author{M.G. Genoni}
\affiliation{Dipartimento di Fisica ``Aldo Pontremoli'', Universit\`a degli Studi di Milano, 20133, Milan, Italy}
\author{M. Barbieri}
\affiliation{Dipartimento di Scienze, Universit\'a degli Studi Roma Tre, Via della Vasca Navale 84, 00146, Rome, Italy}
\affiliation{Istituto Nazionale di Ottica (INO-CNR), L.go E. Fermi 6, I-50125 Firenze, Italy}

\date{\today}

\begin{abstract}

Frequency correlations are a versatile and powerful tool which can be exploited to perform spectral analysis of objects whose direct measurement might be unfeasible. This is achieved through a so-called ghost spectrometer, that can be implemented with quantum and classical resources alike. While there are some known advantages associated to either choice, an analysis of their metrological capabilities has not yet been performed. Here we report on the metrological comparison between a quantum and a classical ghost spectrometer. We perform the estimation of the transmittivity of a bandpass filter using frequency-entangled photon pairs. Our results show that a quantum advantage is achievable, depending on the values of the transmittivity and on the number of frequency modes analyzed. \\

DOI: \url{10.1103/PhysRevA.00.003500}

\end{abstract}

\maketitle

\section{INTRODUCTION} 
Accessing hardly reachable objects with light, while maintaining the possibility of performing detailed analysis at the output is a well-known conundrum, yet a task with substantial relevance.  Correlations in multimode light constitute a widely explored way to circumvent technical limitations. These schemes rely on two sets of correlated modes, which are employed to illuminate the object on one side, and the analysis apparatus on the other. Although the light that has actually encountered the object is not directly analysed, the presence of correlations allows to extract the information. When applied to spatial analysis, this technique is able to produce an image even if the object is physically removed from the detection line, hence the name ghost imaging (GI)~\cite{pittman95pra}.  

Although GI was initially considered as a quantum effect produced in parametric down conversion~\cite{pittman95pra}, many results and features can be replicated by multimode classical thermal emission~\cite{bennink02prl,gatti04prl,valencia05prl}. This has lead to an intense activity focused on extending applications towards genuine remote imaging~\cite{zhao12apl}, as well as extensions to the spectral~\cite{Amiot:18,janassek18appsci} and polarization domains~\cite{lemos14nat}. Thermal GI requires less demanding equipment than its quantum counterpart, and it typically allows to achieve superior brightness. There exist, however, aspects of the quantum technique that can not be replicated with classical light, in particular when inspecting the optical resolution~\cite{bennink04prl}, and the signal-to-noise ratio of the image~\cite{sullivan10pra}. In this article, we discuss quantitative differences of the quantum and classical scheme in the light of a different paradigm in quantum metrology, viz. quantum parameter estimation~\cite{giovannetti04sci, giovannetti06prl, paris09ijqi,PerspectiveMultiPar}.

We discuss the capability of a GI system in estimating the image of an object, considered as a collection of transmission parameters. The presence of quantum correlations is well known to deliver sub-shot noise measurement of intensity~\cite{jakeman86oc,heidmann87prl,mertz90prl,jdrkiewicz04prl,blanchet08prl,brida09prl,iskhakov16ol},
and these can lead to quantum-enhanced applications~\cite{brida10natphot,pirandola11prl,agafonov11ol,clark12oe,lopaeva13prl,lawrie13prl,pooser15opt,pooser16acs,moreau17scirep,samantaray17lsa}.
In particular, the authors of Ref.~\cite{D_Auria_2006} discussed the use of quantum light for the measurement of a single transmittance, while the authors of Ref.~\cite{losero18scirep} demonstrated how this task benefits from adopting quantum correlations. We build on these results to discuss resource counting in quantum GI in comparison with its classical counterpart at fixed energy.

\section{EXPERIMENTAL QUANTUM GI IN THE SPECTRAL DOMAIN}

In our approach, the object to be imaged - be it a genuine spatial image or a spectral profile, is modeled as a collection of $K$ values of transmittivity $T_k$, $1\leq k \leq K$, each associated to a  mode. Our aim is then to estimate these values. 

The quantum technique to implement GI uses the correlations between single photons produced in spontaneous parametric down-conversion (SPDC): a single incoming pump photon creates a pair of photons strongly correlated in their emission modes \cite{review}. In the experiment, the spectral profile is  conveniently discretized, so that the effective number of correlated mode pairs is equal to $K$.

\begin{figure}[h!]
    \includegraphics[width =\columnwidth]{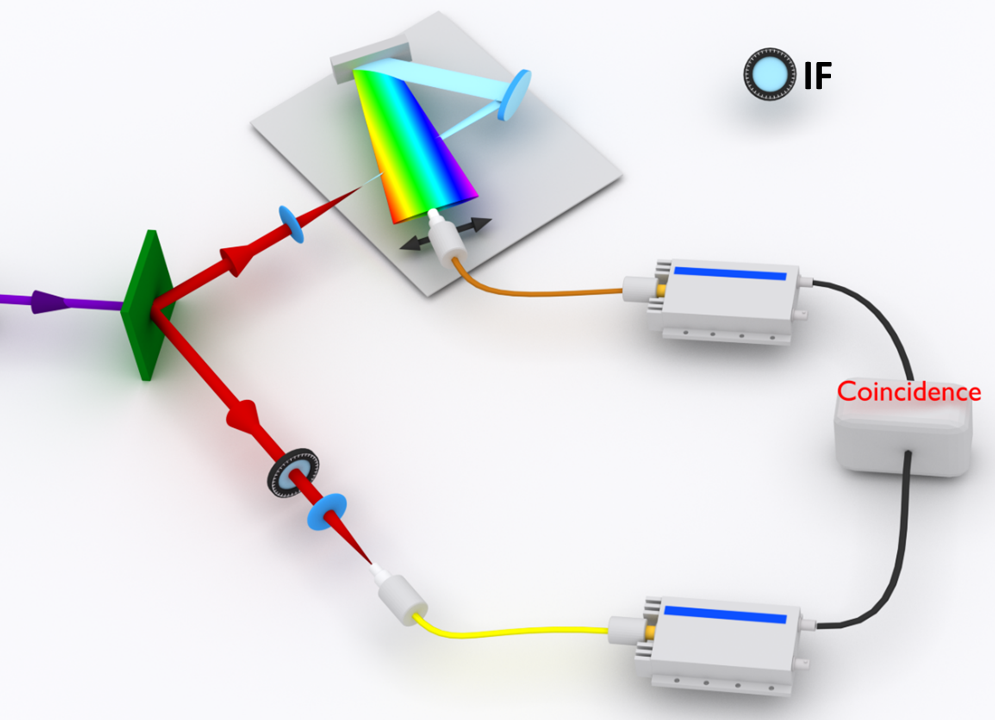}
    \caption{Experimental setup. A 30 mW CW diode laser at 405 nm pumps a 3-mm barium borate (BBO) crystal cut for non-collinear Type I phase matching, producing degenerate photon pairs at 810 nm through SPDC. One photon is then sent through an interference filter with FWHM 7.3 nm, and detected with a bucket detector. Frequency detection is performed on the second photon: this is achieved through  a 1200-lines/mm grating and a collection lens with focal length $f=30$ cm. A multimode fiber on a translation stage allows for a complete measurement of the spectral range under investigation. A collimator, integral to the fiber mount, assists the photons collection.  } 
    \label{fig:setup}
\end{figure}

All the modes in arm 1 arrive on the object, and are then measured by a mode-insensitive bucket detector. Due to the correlations in the pair production process, the analysis of the correlated photon in coincidence with the bucket detector provides information about what has occurred to its twin. This is the experimental approach we followed in our investigation, but, differently from the most frequent case, we explored the spectral domain. The photon reaching the frequency-independent bucket detector passes through a spectral object, an interference filter with full-width at half-maximum of 7.3 nm. The second photon was analyzed using a spectrometer, as described in Fig.\ref{fig:setup}. 

\begin{figure*}[t!!!]
   \includegraphics[width =\textwidth]{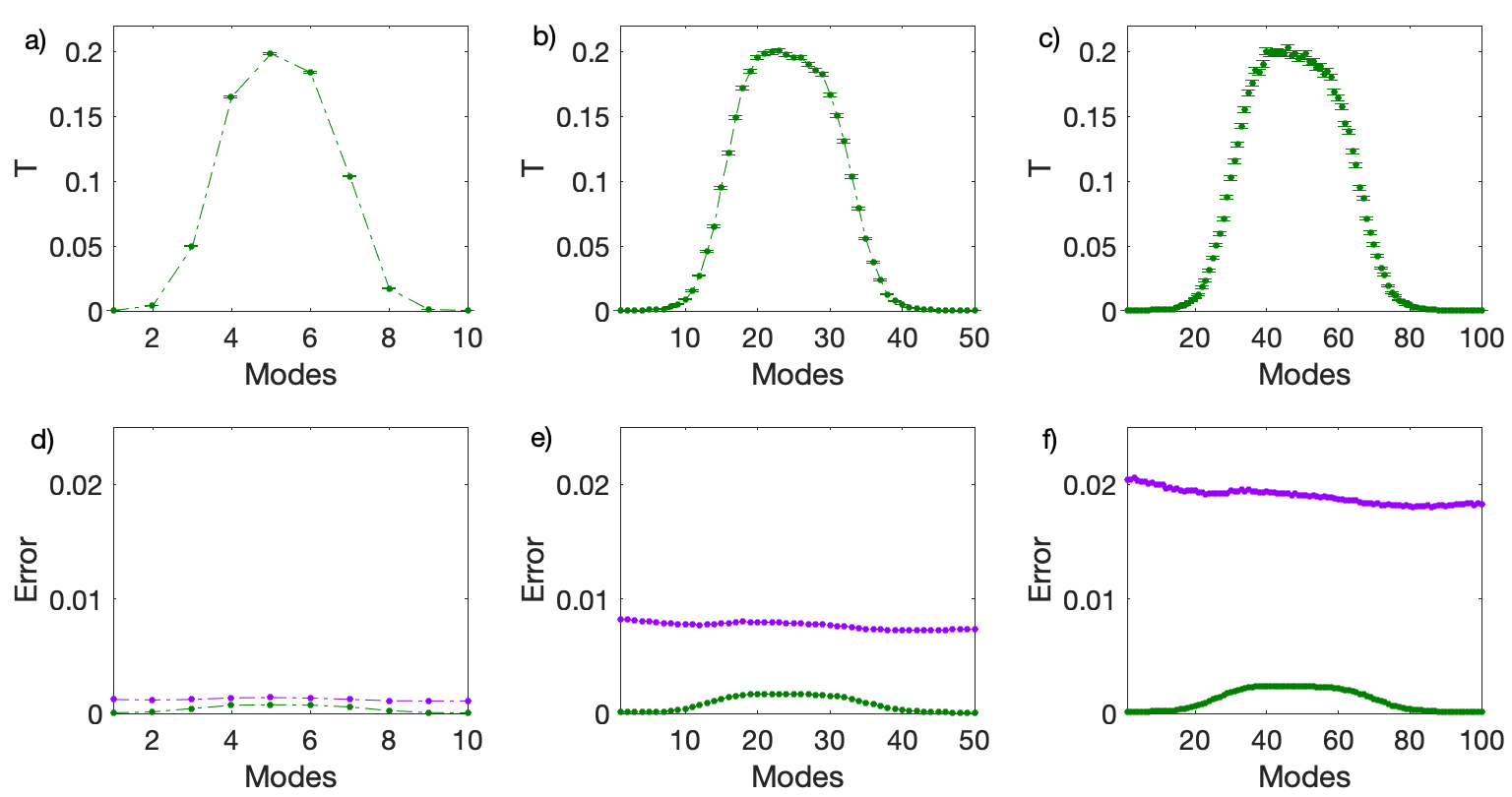}
   \caption{Transmittivities evaluated as the Klyshko efficiency for a) $K$=10, b) $K$=50, and c) $K$= 100 modes. Data are collected for the spectral resolution of panel c) in a 5 s window with a rate of 1100 coincidences/s at the maximum. The values of transmittivity are estimated as the coincidence-to-single count ratio $C_k/N_k$. The points in panel c) are obtained directly  from the measured counts, while for the points in panels a and b, the reduced resolution is mimicked by summing the collected signals over multiple modes. The observed profile follows closely the fourth-order super-Gaussian, previously measured by a spectrophotometer~\cite{bbo}, with some deviations which can be ascribed to other optical components in the arm. The experimental estimation errors for the quantum strategy are obtained by means of Eq.\eqref{propag_quant}, and are shown in green in panel d) for $K$= 10, e) $K$=50, and f) $K$=100 modes. These correspond to the error bars in the panels (a) to (c). The classical uncertainties (purple points) are evaluated via the error propagation in Eq.\eqref{propagazione} and the described results of our calculation. We considered $N_{tot}=80000$ per mode at the highest resolution, a value that accounts for the lost events due to the limited efficiency $\eta=0.35$.
   }
   \label{fig:Tcurve}
\end{figure*}

We reconstruct the spectral profile of arm 1, including the filter, optical elements, and detector, by scanning the output of the spectrometer in the analysis arm 2. This profile will be dictated mostly by that of the filter. We collect a series of $N_s=100$ points, with a resolution of 0.33 nm, estimated by comparing the obtained profile with the one measured in Ref. \cite{bbo} by means of a conventional spectrophotometer. We can then define the spectral modes as $K=N_s/j$, with $j= 1--100$, so that each single mode $k$ is obtained by regrouping $j$ measured points. Different spectral resolutions were thus achieved by summing  the number of coincidences measured for these groups of points, and similarly for the single counts.
The transmission $T_k$, considering the spectral object as well as the intrinsic loss of the arm, is calculated as a Klyshko efficiency \cite{klyshko80sjqe}$T_k=C_k/N_k$, with $C_k$ being the coincidence counts and $N_k$ the total counts of the spectrometer detector for the $k$-th mode; this allows to obtain an estimation of each $T_k$, independently of the other. In Figs.\ref{fig:Tcurve} (a) to (c) we report the obtained transmittivities $T_k$ for $K= 10,50, 100$.  

The uncertainty on the transmittivity is calculated by considering that $N_k$ events have been collected, a fraction $C_k$ of which lead to a coincidence. Thus, $N_k$ is considered as fixed, while $C_k$ is a binomial variable, with variance $C_k(1-T_k)$~\cite{losero18scirep}; error propagation leads to: 
\begin{equation}
\Delta^2 T_k = T_k(1-T_k)/N_k.
\label{propag_quant}
\end{equation}
The adoption of the Klyshko method makes the estimation of $T_k$ and its error independent on the efficiency $\eta$ of the detection arm, within our single-photon approximation. However, proper resource counting needs to include those events discarded due to $\eta<1$: for each value of $T_k$, these are estimated as $N_{\rm tot}=\langle N_k \rangle /\eta$, using the average number of events on all modes. In our experiment we estimated $\eta=0.35$ as the average efficiency of the frequency bins by a modified Klyshko method that takes into account the multimode detection on arm 1. This consists in evaluating $\eta$ by taking the sum of the coincidences across all frequency detection modes, and dividing it by the total counts of the bucket detector.

\begin{figure*}[t!!]
   \includegraphics[width =\textwidth]{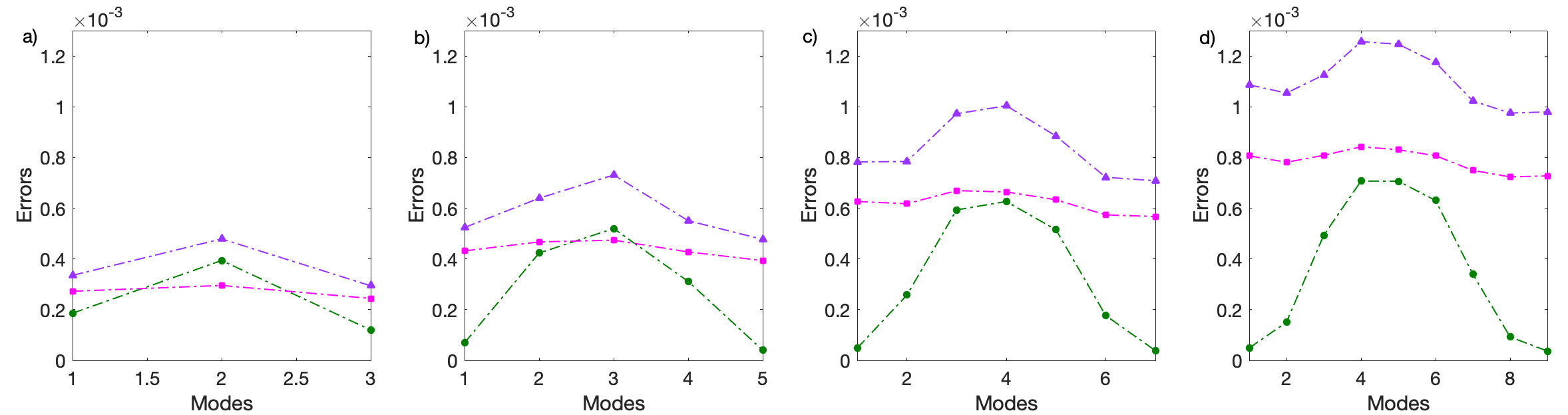}
   \caption{Comparison between the measured errors for the quantum strategy (green circles), and the estimated errors for the classical one obtained by propagation (purple triangles) and by the CRB (pink squares), at fixed number of cumulative resources $N_{\rm tot}$ for a) $K$ = 3, b) $K$ = 5, c) $K$ = 7, and d) $K$ = 9 modes.}
   \label{fig:fishers}
\end{figure*}

\section{BENCHMARKING THERMAL GI}

We derive the equivalent strategy based on multiple thermal states, as our classical benchmark. We assume we can make use of a collection of independent modes, with the same structure as our quantum source, each presenting thermal population. This multi-thermal emission is split on a 50:50 beam splitter, with the measuring apparatus performing essentially the same operation as above: one half of the beam reaches the object and then the bucket detector, the other half the analysis apparatus. Looking at the cross-correlation between the two detection signals, one observes a value of second-order correlation $g^{(2)}(0)>1$  if the two are correlated, and $g^{(2)}(0)=1$ otherwise. The value of the second-order correlation will depend on the transmission profile, thus making it  possible to obtain information on the object. Since $g^{(2)}(0)\geq 1$ for classical light, the visibility of our signal is decreased with respect to the quantum case~\cite{gatti04prl,valencia05prl,bennink04prl,sullivan10pra}. Notably, this mechanism can not be replicated by means of coherent states, since no intensity correlations emerge when these are divided on a beam splitter.


A multimode thermal state can be written in the diagonal form in the Fock basis as \cite{GaussFOP}
\begin{align}
{\boldsymbol \nu}_{n_{\rm th}} &= \bigotimes_{k=1}^K \sum_{m=0}^{\infty} p_{\rm th}(m | n_{\rm th}) \, |m\rangle_k \, {}_k\langle m| \,,
\end{align}
where the thermal profile is given by the photon-number probability 
\begin{align}
    p_{\rm th}( m | n_{\rm th})&= \frac{1}{n_{\rm th} + 1} \left(\frac{n_{\rm th}}{n_{\rm th}+1} \right)^m  \,.
\end{align}
Each thermal mode is taken to have mean photon number $n_{th}$, so that, on average, $\bar{n}=n_{th}/2$ photons per mode reach the object, and allow for $M$ repetitions of the measurement such that $\bar{n}M = N_{\rm tot}$, to compare strategies with the same number of total resources. Near-optimal working conditions are for $\bar{n} \sim 1$, as verified numerically. We should remark that this comparison is carried out against the post-selected scheme of quantum metrology.

The thermal state is split in a balanced beam-splitter and, as in the experiment described in Fig.\ref{fig:setup}, one half of the beam reaches the filter and then the bucket detector, while the other half reaches the analysis apparatus. The key quantity needed to evaluate the classical benchmark for ghost imaging, that is the error that one would obtain in estimating the same set of transmittivities $\{T_k \}$ describing arm 1, is the joint conditional probability $P_k(n_1, n_2 | \{T_i\})$ of detecting $n_1$ photons in the bucket detector, and $n_2$ photons in the mode $k$ via the detector placed after the frequency analyzer. 

We start by considering the single-mode case, where our object is thus described by a beam-splitter with transmittivity $T_k$ that couples the photons in the arm 1 to a virtual mode $``0''$ prepared in a vacuum. By assuming to have control on this virtual mode and thus to detect photons also in the corresponding output port, the corresponding conditional probability of detecting respectively $\{ n_1, n_2, n_0\}$ in the three detectors reads
\begin{align}
p_k(n_1,& n_2,n_0 | T_k )= \binom{n_1+n_2+n_0}{n_1+n_0} \binom{n_1+n_0}{n_0}  \nonumber \\
& \tiny{\times}\,\, p_{\rm th}(n_1+n_2+n_0 | n_{\rm th})\left(\frac{1}{2}\right)^{n_1+n_2+n_0} T_k^{n_1} (1-T_k)^{n_0} \,.
\end{align}
To obtain the correct conditional probability corresponding to the output of the two detectors in the actual experiment, where one does not have control on the virtual mode, we have to trace out this subsystem by averaging over all the possible values of $n_0$, obtaining the marginal probability  
\begin{align}
p_k(n_1,n_2 | T_k ) = \sum_{n_0=0}^{\infty}  p_k(n_1,n_2,n_0 | T_k ) \,.   
\label{eq:singlemode}
\end{align}

As expected, by averaging also over the detector output $n_2$ one obtains the photon-number statistics of a thermal state with $T_k n_{\rm th}/2$ photons, i.e.
\begin{align}
\sum_{n_2=0}^{\infty} p_k(n_1,n_2 | T_k ) = p_{\rm th}(n_1 | T_k \bar{n}) \,.
\label{eq:average}
\end{align}

In the multimode scenario, we have to consider two contributions to the detection. When the detector on arm 2 is set to observe mode $k$, the bucket detector can receive photons originating from the correlated mode on arm 1, or from the other uncorrelated modes. In the first case, the detection probability has the expression for $p_k(n_1,n_2|T_k)$ calculated in Eq.(\ref{eq:singlemode}) for the single-mode scenario, while in the second case there will be present multi-thermal noise, with each mode contributing with its thermal statistics $p_{\rm th}(n_1 | T_k \bar{n})$ in Eq.~(\ref{eq:average}). Consequently in the complete conditional probability $P_k(n_1,n_2|T_k)$ for the multimode case, we have to account for the possibility of generating the photons $n_1$ in arm 1 from all these modes: of the observed $n_1$ photons, $\nu_k$ actually originate from mode $k$, and $n_1-\nu_k$ from the others, parted among the remaining modes. The overall probability corresponds to taking the discrete convolution of those for the individual modes:
\begin{align}
    P_k(n_1,n_2 | \{T_i \})= \sum_{\substack{\{\nu_i\}: \\ \sum_i \nu_i = n_1}} 
    p_k(\nu_k,n_2 | T_k )\prod_{j\neq k} p_{th}(\nu_j|T_j \bar{n}) \,, 
    \label{thehorror}
\end{align}
where the sum is indeed taken over all the possible set $\{\nu_i \}$ of photons on the $K$ modes hitting the bucket detector.

The conventional measurement estimates the correlation $C_{12}^{(k)}=\langle n_1 n_2\rangle_k$, from which the transmittivities $T_k$ can be inferred (notice that each trasmittivity $T_k$ will be estimated separately by selecting only the clicks of the second detector corresponding to the $k$-th frequency). The corresponding uncertainties can thus be evaluated via error propagation as
\begin{equation}
\label{propagazione}
    \Delta^2T_k^{(c)} = \frac{1}{\left(dC_{12}^{(k)}/dT_k\right)^2} \frac{\Delta^2C_{12}^{(k)}}{M},
\end{equation}
where 
$\Delta^2C_{12}^{k} = \langle n_1^2 n_2^2\rangle_k - \langle n_1 n_2\rangle_k^2$. However the evaluation of $C_{12}^{(k)}$ and $\Delta^2C_{12}^{(k)}$ directly from Eq. (\ref{thehorror}) is computationally demanding. We thus adopted an approach based on the moment generating functions~\cite{ROUSSAS2014163}, defined as
\begin{align}
    G_k(x,y | \{ T_i \} ) = \sum_{n_1,n_2=0}^{\infty}P_k(n_1,n_2 | \{T_i \} )e^{n_1x+n_2y} \,.
\label{Eq:gener}
\end{align}
In fact, by exploiting these objects, the moments of the distributions are found as:
\begin{equation}
    \langle n_1^\alpha n_2^\beta \rangle_k = \partial_x^\alpha \partial_y^\beta G_k(x,y | \{T_i \} )\vert_{x=0,y=0}. 
\end{equation}
The key property we exploit to evaluate $G_k(x,y | \{ T_i \} )$ is that for probabilities based on a convolution such as Eq. (\ref{thehorror}), the total generating function is readily found as the product of the individual functions
\begin{equation}
    G_k(x,y | \{T_i\})=g_k(x,y|T_k)\prod_{j\neq k}g_{th}(x|T_j \bar{n}) \,,
\label{Eq:gener2}
\end{equation}
that can be readily evaluated via the formulas
\begin{align}
    g_k(x,y | T_k) &= \sum_{n_1,n_2=0}^{\infty} p_k(n_1,n_2 | T_k ) e^{n_1x+n_2y} \,, \\
    &= [1 + n_{th} (1 + T_k -e^y - T_k e^x) ]^{-1}\,, \\
    g_{th}(x|n_{th}) &= \sum_{n_1=0}^{\infty} p_{th}(n_1 |n_{th}) e^{n_1 x} \, \\
    &= (1 + n_{th} - n_{th} e^x )^{-1} \,.
\end{align}
By exploiting the results of our calculation it is possible to obtain computable forms for the uncertainties in \eq{propagazione}, this is also true for large values of the number of modes $K$. We remark that the comparison between the estimation errors for the quantum apparatus described in the previous section and the the classical strategy based on thermal light is conducted by allowing the classical strategy to employ also those resources that were wasted due to the loss in the quantum apparatus, as quantified by the parameter $\eta$.   

This classical strategy is inspired by the standard measurement carried out for ghost imaging; furthermore, since there is no coherence among the different photon number states, the choice of the observable is bound to be optimal. However, the estimator $C_{12}$ may be not: a more suitable choice $f(n_1,n_2)$, while based on the same observable, may lead to improved uncertainties. On the other hand, finding an explicit expression, due to the dissipative nature of the transmission process, is not immediate, as we can not apply the standard machinery for unitary parameters. Anyhow, the ultimate limit on the error in the estimation of each trasmittivity $T_k$ via the experimental setup described above is given by the Cram\'er-Rao bound (CRB) \cite{paris09ijqi}:
\begin{align}
    \Delta T_k^2 \geq \frac{1}{M F_k} \,,
\label{Eq:CRB}
\end{align}
where $M$ is the number of repetitions of the experiments, while 
\begin{align}
    F_k = \sum_{n_1,n_2=0}^{\infty} P_k(n_1,n_2 | \{T_i \}) \left( \frac{\partial  }{\partial T_k} \log P_k (n_1,n_2 | \{T_i \} ) \right)^2
\label{fisher}
\end{align}
denotes the Fisher information corresponding to the estimation of the parameter $T_k$. While this does not apply strictly to a genuine multiparameter estimation of all $\{T_k\}$ \cite{PerspectiveMultiPar}, it still sets a lower bound to the attainable error in the general case (accounting for statistical correlations among transmittivities can only decrease the available information).

For large number of modes $K$ the evaluation of the probabilities in \eq{fisher} is highly demanding. Hence, we employed the approach described in the previous paragraph to evaluate the Fisher information, and thus the corresponding CRB, only numerically and for a small number of modes $K$, by exploiting the exact relationship between the Fisher information and the Hellinger distance \cite{Hellinger}

\begin{align}
    F_k = \lim_{\epsilon \rightarrow 0} \frac{4 \,\left(\mathcal{D}_H[ P_k(n_1,n_2 | \{T_i \}) , P_k^{(\epsilon)}(n_1,n_2 | \{T_i \})]\right)^2}{\epsilon^2} \,,
\end{align} 
where $P_k^{(\epsilon)}$ is obtained from $P_k$ by replacing the $k$-th trasmittivity with $T_k + \epsilon$, and where we defined the Hellinger distance between two probability distributions as
\begin{align}
    \mathcal{D}_H[p_1(x), p_2(x)] = \sqrt{ \sum_x \left( \sqrt{p_1(x)} - \sqrt{p_2(x)} \right)^2 } \,.
\end{align}

\section{DISCUSSION}
The errors evaluated for quantum and classical strategies, as reported in Eqs. (\ref{propag_quant}) and (\ref{propagazione}), respectively, are reported in Figs. \ref{fig:Tcurve}(d) to \ref{fig:Tcurve}(f) for $K=10,50, 100$. This shows that for the conventional estimators, the quantum strategy, although lossy, always outperforms the classical one, and that the enhancement increases with the number of modes, when the transmittivities are estimated individually for each mode.

The comparison between the errors evaluated for the quantum strategy [\eq{propag_quant}], for the classical one through propagation [\eq{propagazione}] and through the CRB [\eq{Eq:CRB}] is reported in Fig. \ref{fig:fishers}, for modes $K= 3,5,7,9$. In more detail, to obtain the results shown in Fig. \ref{fig:fishers}, we evaluated numerically the probability distributions $P_k$ and $P_k^{(\epsilon)}$ via \eq{thehorror} with $\epsilon = 10^{-7}$, numerically checking that the quantity
\begin{equation}
    \tilde{F}_k = \frac{4 \,\left(\mathcal{D}_H[ P_k(n_1,n_2 | \{T_i \}) , P_k^{(\epsilon)}(n_1,n_2 | \{T_i \})]\right)^2}{\epsilon^2}
\end{equation}
is stable by further decreasing the value of $\epsilon$, such that one can safely assume that $F_k \approx \tilde{F}_k$.

The results show that quantum light does not provide an advantage unconditionally. In fact, when only a few modes are considered, an optimal classical estimator can outperform the quantum strategy using the same number of resources, especially at higher transmittivities. When the number of modes is increased, however, the quantum advantage is recovered, hence the quantum estimation becomes the preferable choice for every transmittivity value considered. It should be noted that, to be optimal, the classical protocol requires $n_{th}\sim 1$, thus prompting comparable experimental difficulties to those of the quantum scenario. This is indeed quite a different regime than the one for conventional thermal ghost imaging~\cite{valencia05prl}.

\section{CONCLUSIONS} 
We investigated in which conditions a ghost imaging setup may provide an enhanced parameter estimation of a collection of transmittivity values representing the imaged object. We illustrated this with an experiment of quantum ghost spectrometry performing the measurement of a bandpass filter. We then compared the measurement precision with that of an analogous classical scheme using thermal modes. Our analysis shows that adopting the quantum strategy can be favorable in specific conditions, dictated by the values of the transmittivities at hand, and by resolution of the modes. The higher the resolution and the lower the transmittivities, the greater the enhancement that can be achieved through quantum estimation, although, it should be emphasized that the details do depend on the entire profile of the transmittivities.

\section*{ACKNOWLEDGMENTS} 
The authors thank M. Guarneri for assistance with the numerical calculations, G. Satta for assistance in the laboratory, and V. Berardi for fruitful discussion. This work was supported by the European Commission (FET-OPEN-RIA STORMYTUNE, Grant Agreement No. 899587), and by the NATO SPS Project HADES - MYP G5839.


\end{document}